\documentclass{article}

\usepackage{arxiv}
\usepackage{amsmath,amssymb,amsfonts}
\usepackage{algorithmic}
\usepackage{graphicx}
\usepackage{textcomp}
\usepackage{xcolor}
\usepackage{multirow}   
\usepackage{graphicx}   
\def\BibTeX{{\rm B\kern-.05em{\sc i\kern-.025em b}\kern-.08em
    T\kern-.1667em\lower.7ex\hbox{E}\kern-.125emX}}
\begin{document}

\title{Intrusion Detection in Internet of Vehicles Using Machine Learning\\
}

\author{
    Hop Le \\
    Department of Computational, Engineering, and Mathematical Sciences\\
    Texas A\&M University- San Antonio\\
    San Antonio, Texas \\
    \texttt{hle014@jaguar.tamu.edu} \\
    \AND
    Izzat Alsmadi\\
    Department of Computational, Engineering, and Mathematical Sciences\\
    Texas A\&M University - San Antonio\\
    \texttt{ialsmadi@tamusa.edu}\\
}
\maketitle
\begin{abstract}
The Internet of Vehicles (IoV) has evolved modern transportation through enhanced connectivity and intelligent systems. However, this increased connectivity introduces critical vulnerabilities, making vehicles susceptible to cyber-attacks such Denial-ofService (DoS) and message spoofing. This project aims to develop a machine learning-based intrusion detection system to classify malicious Controller Area network (CAN) bus traffic using the CiCIoV2024 benchmark dataset. We analyzed various attack patterns including DoS and spoofing attacks targeting critical vehicle parameters such as Spoofing-GAS - gas pedal position, Spoofing-RPM, Spoofing-Speed, and Spoofing-Steering\_Wheel. Our initial findings confirm a multi-class classification problem with a clear structural difference between attack types and benign data, providing a strong foundation for machine learning models.
\end{abstract}

\section{Introduction}
The Internet of Vehicles (IoV) concepts are transforming auto safety and convenience. Modern vehicles contain numerous Electronic Control Units (ECUs) that communicate through the Controller Area Network (CAN) bus protocol. While this connectivity enables advanced features such as adaptive cruise control, lane-keeping assistance, and vehicle-to-vehicle communication, the CAN bus, the backbone of in-vehicle communication, lacks inherent security features like authentication or encryption. This exposes vehicles to cyber threats that were previously non-existent. This makes it a prime target for cyber-attacks. 

Recent studies have demonstrated that attackers can exploit vulnerabilities in vehicle communication systems to compromise safety-critical functions\cite{k6}. Spoofing attacks can manipulate sensor readings, potentially causing incorrect vehicles behavior, while Denial of Service (DoS) attacks can disrupt communication between ECUs, leading to system failures. A successful attack can range from privacy breaches to life-threatening situations like taking control of a vehicle's brakes or steering\cite{k6}. 

Intrusion Detection Systems (IDS) are therefore critical for modern vehicle security. Traditional signature-based IDS are ineffective against zero-day and sophisticated attacks\cite{k6}. Machine learning offers a promising solution by learning the normal behavior of CAN bus traffic and flagging significant deviations as potential threats\cite{k6}.

\section{Dataset}
The project uses the CICIoV2024 dataset\cite{k7} showed in TABLE 1, a benchmark dataset for evaluating IoV security solutions

\begin{table}[htbp]
    \centering
    \caption{Dataset Statistics}
    \begin{tabular}{|l|c|}
        \hline
        \textbf{Dataset} & \textbf{Records} \\
        \hline
        decimal\_benign & 1,223,737 \\
        decimal\_DoS & 74,663 \\
        decimal\_spoofing-GAS & 9,991 \\
        decimal\_spoofing-RPM & 54,900 \\
        decimal\_spoofing-SPEED & 24,951 \\
        decimal\_spoofing-STEERING\_WHEEL & 19,977 \\
        \hline
        \textbf{Total Combined} & \textbf{1,408,219} \\
        \textbf{Total Features} & \textbf{12} \\
        \hline
    \end{tabular}
    \label{tab:dataset_stats}
\end{table}

\section{Related Codes}
In the context of CAN bus security, recent implementations have demonstrated the efficacy of aggregating benign traffic with specific attack vectors, such as Denial of Service (DoS) and spoofing attacks, to create comprehensive training sets in this "Intrusion Detection System"\cite{k1}. A critical component of this workflow is the schema standardization process, where iterative whitespace stripping from column headers prior to concatenation ensures data integrity across disparate dataframes. However, effective analysis relies heavily on the granular sanitation of categorical variables; omitting whitespace cleaning for label and category columns results in the erroneous bifurcation of identical classes, undermining visualization accuracy and subsequent model training in "Intrusion Detection System"\cite{k1}.

Moving beyond automotive protocols to general network security in "Cybersec25-challenge-1"\cite{k3}, the detection of ARP spoofing via PCAP exports requires distinct preprocessing strategies, particularly when handling non-finite numerical data. Successful approaches have employed maximum finite value replacement for infinite rates to ensure compatibility with standard scaling techniques, a prerequisite for distance-based unsupervised algorithms like Isolation Forest and One-Class SVM\cite{k3}. By integrating feature engineering steps that capture flow variability, such as packet size ranges, and employing variance thresholding to reduce dimensionality, these pipelines effectively isolate anomalies without the need for extensive labeled references. Conversely, in the domain of IoT security, the massive scale of datasets like CICIoT2023 in "CICIoT2023 prediction"\cite{k2}, demands resource-efficient strategies. Validating preprocessing logic on single data partitions before scaling to the full dataset allows for the identification of severe class imbalances across dozens of attack types\cite{k2}. To mitigate this, practitioners often map detailed attack signatures into broader macro-categories and employ Random Under-Sampling (RUS); however, while RUS reduces computational load, it risks discarding valuable forensic data, suggesting that class weighting may offer superior performance for production models\cite{k2}.

Hybrid methodologies in "IoT Network Vulnerabilities"\cite{k4} that bridge unsupervised and supervised learning have shown particular promise in characterizing complex IoT threat landscapes. By utilizing K-Means clustering to partition traffic into semantically meaningful groups—such as floods or botnets—analysts can validate latent behaviors prior to supervised classification\cite{k4}. While ensemble methods like XGBoost and Random Forest achieve near-perfect accuracy on dominant classes within these pipelines, they frequently exhibit critical deficiencies in detecting rare, low-frequency attacks like injections, necessitating the integration of synthetic data generation techniques (e.g., SMOTE) or specialized anomaly detection modules. Finally, large-scale ingestion pipelines utilizing pattern matching, in "ciciot2023"\cite{k5}, to concatenate millions of traffic records have confirmed that specific flow metrics, such as Inter-Arrival Time (IAT), serve as primary discriminators for tree-based classifiers\cite{k5}. Nevertheless, the failure of linear models like Logistic Regression in these environments highlights the non-linear nature of network attack traffic and underscores the absolute necessity of rigorous data hygiene—specifically the standardization of feature names and categorical labels—to ensure model convergence and interpretability across all utilized algorithms\cite{k5}.

\section{Related Work}
\label{sec:Related Work}
The introduction of the CICIoT2023 dataset\cite{r1} provided a necessary stress test for traditional machine learning algorithms, establishing a consensus across multiple studies regarding classifier efficacy. Both the foundational analysis\cite{r1} and the subsequent comparative evaluation\cite{r2} agree that while binary detection is trivial, the granular classification of 33 distinct attack vectors exposes the limitations of linear models; Logistic Regression and SVMs consistently fail in these high-dimensional environments. Conversely, Random Forest emerges as the dominant architecture in both studies, handling the feature complexity with superior F1-scores. Expanding on this baseline, the research in\cite{r3} demonstrates that this complexity can be managed not just by model selection, but by data reduction. By applying Gaining-Sharing Knowledge (GSK) clustering, they achieved a 62\% reduction in dimensionality while allowing a Multilayer Perceptron (MLP) to outperform standard AutoEncoders, suggesting that efficient feature engineering can bridge the gap between heavy ensemble models and the need for speed.

The "Perfect Score"\cite{r7} Anomaly in IoV Classification. A critical synthesis of the CICIoV2024 literature reveals a striking discrepancy between the dataset's inherent difficulty and reported model performance. The dataset creators\cite{r7} explicitly note that while DoS attacks are distinct, spoofing attacks—particularly Speed Spoofing—are difficult to classify due to their resemblance to benign traffic. However, subsequent studies utilizing advanced sampling techniques appear to trivialize this complexity. Paper \cite{r8} employs hybrid sampling (downsampling/oversampling) and reports perfect 1.0 precision/recall scores across all categories using HistGradientBoosting. Similarly, Paper \cite{r9} applies Random Under-Sampling (RUS) to balance the data and also reports perfect 1.0 scores with XGBoost, noting a drastic reduction in training time to just 5.6 seconds. While these results highlight the power of gradient boosting and balanced data, the leap from "challenging dataset"\cite{r7} to "perfect classification"\cite{r8, r9} warrants scrutiny regarding potential data leakage or the possibility that synthetic balancing has simplified the decision boundaries too aggressively.

Deep Learning Performance and the Risk of Overfitting. The application of Deep Neural Networks (DNN) to these datasets highlights a tension between raw metrics and model generalizability. Paper \cite{r4} reports an exceptionally high accuracy of 99.8\% on the CICIoT2024 dataset, achieving perfect 1.0 scores for DoS and Spoofing categories. While impressive, such metrics often signal a model that has memorized the training distribution rather than learning robust features. This skepticism is validated by the comparative analysis in \cite{r10}, where researchers achieved similar 99\% accuracy using CNNs and LSTMs on IoV data but candidly concluded that such scores "most presumably can be an illustration of over-training." This reflection is vital; it suggests that the field may be reaching a saturation point where "perfect" metrics on static datasets no longer correlate with real-world robustness, necessitating a shift toward adversarial testing.

Data Availability and Edge Constraints Two systematic reviews address the macro-level hurdles preventing the deployment of these theoretical models. Paper \cite{r5} identifies the root cause of stagnation as a lack of open, qualitative data, arguing that without mandatory incident reporting and adherence to FAIR principles, cyber risk modeling remains speculative. Paper \cite{r6} complements this by noting that even when data is available, the computational cost of the high-performing models (like the DNNs and Ensembles mentioned above) makes them unsuitable for Edge IoT devices. The authors advocate for a paradigm shift toward decentralized architectures, utilizing Federated Learning and Blockchain to resolve the conflict between the need for heavy security models and the resource constraints of the IoT edge.

\section{Data Analytics}
\subsection{Data Load}
The workflow begins by importing necessary libraries (pandas, numpy, matplotlib, seaborn) and initializing the plotting style. The data ingestion process involves loading six specific CSV files representing different traffic categories, 'decimal\_benign', 'decimal\_DoS', and four variations of 'decimal\_spoofing' (GAS, RPM, SPEED, STEERING\_WHEEL). These files are concatenated into a unified DataFrame, resulting in a total of 1408,219 records and 12 features.

\begin{table}[htbp]
    \centering
    \caption{Category Distribution}
    \begin{tabular}{|l|c|c|}
        \hline
        \textbf{Category} & \textbf{Count} & \textbf{Percentage} \\
        \hline
        BENIGN & 1,223,737 & 86.90\% \\
        SPOOFING & 109,819 & 7.80\% \\
        DoS & 74,663 & 5.30\% \\
        \hline
    \end{tabular}
    \label{tab:category_distribution}
\end{table}

\begin{table}[htbp]
    \centering
    \caption{Data Types}
    \begin{tabular}{|l|c|}
        \hline
        \textbf{Column} & \textbf{Data Type} \\
        \hline
        ID & int64 \\
        DATA\_0 & int64 \\
        DATA\_1 & int64 \\
        DATA\_2 & int64 \\
        DATA\_3 & int64 \\
        DATA\_4 & int64 \\
        DATA\_5 & int64 \\
        DATA\_6 & int64 \\
        DATA\_7 & int64 \\
        label & object \\
        category & object \\
        specific\_class & object \\
        \hline
    \end{tabular}
    \label{tab:data_types}
\end{table}

\subsection{Data Preprocessing}

\begin{table*}[htbp]
    \centering
    \caption{First 5 Rows of the Dataset}
    \resizebox{\textwidth}{!}{
        \begin{tabular}{|c|c|c|c|c|c|c|c|c|c|c|c|c|}
            \hline
            \textbf{Index} & \textbf{ID} & \textbf{DATA\_0} & \textbf{DATA\_1} & \textbf{DATA\_2} & \textbf{DATA\_3} & \textbf{DATA\_4} & \textbf{DATA\_5} & \textbf{DATA\_6} & \textbf{DATA\_7} & \textbf{label} & \textbf{category} & \textbf{specific\_class} \\
            \hline
            0 & 65   & 96  & 0   & 0   & 0   & 0   & 0   & 0   & 0   & BENIGN & BENIGN & BENIGN \\
            1 & 1068 & 132 & 131 & 6   & 0   & 0   & 0   & 0   & 0   & BENIGN & BENIGN & BENIGN \\
            2 & 535  & 127 & 255 & 127 & 255 & 127 & 255 & 127 & 255 & BENIGN & BENIGN & BENIGN \\
            3 & 131  & 152 & 24  & 0   & 0   & 0   & 0   & 0   & 0   & BENIGN & BENIGN & BENIGN \\
            4 & 936  & 103 & 91  & 6   & 0   & 0   & 0   & 0   & 0   & BENIGN & BENIGN & BENIGN \\
            \hline
        \end{tabular}
    }
    \label{tab:first_five_rows}
\end{table*}

\begin{table*}[htbp]
    \centering
    \caption{Basic Statistics}
    \resizebox{\textwidth}{!}{
        \begin{tabular}{|l|c|c|c|c|c|c|c|c|c|}
            \hline
            \textbf{Statistic} & \textbf{ID} & \textbf{DATA\_0} & \textbf{DATA\_1} & \textbf{DATA\_2} & \textbf{DATA\_3} & \textbf{DATA\_4} & \textbf{DATA\_5} & \textbf{DATA\_6} & \textbf{DATA\_7} \\
            \hline
            count & 1,408,219 & 1,408,219 & 1,408,219 & 1,408,219 & 1,408,219 & 1,408,219 & 1,408,219 & 1,408,219 & 1,408,219 \\
            mean & 537.21 & 71.09 & 69.99 & 55.01 & 57.45 & 45.29 & 53.88 & 71.75 & 60.27 \\
            std & 322.48 & 88.98 & 95.58 & 72.77 & 90.32 & 64.46 & 94.34 & 101.69 & 99.97 \\
            min & 65.00 & 0.00 & 0.00 & 0.00 & 0.00 & 0.00 & 0.00 & 0.00 & 0.00 \\
            25\% & 357.00 & 0.00 & 0.00 & 0.00 & 0.00 & 0.00 & 0.00 & 0.00 & 0.00 \\
            50\% & 516.00 & 16.00 & 12.00 & 13.00 & 0.00 & 6.00 & 0.00 & 0.00 & 0.00 \\
            75\% & 578.00 & 127.00 & 128.00 & 125.00 & 92.00 & 86.00 & 63.00 & 138.00 & 80.00 \\
            max & 1,438.00 & 255.00 & 255.00 & 255.00 & 255.00 & 255.00 & 255.00 & 255.00 & 255.00 \\
            \hline
        \end{tabular}
    }
    \label{tab:basic_statistics}
\end{table*}

\textit{Missing Values:} An initial quality assessment confirmed that there were zero missing values in the raw dataset. However, a robust handling strategy was implemented to fill numeric columns with the median and categorical columns with the mode if nulls were encountered during processing.
\textit{Duplicates:} A high duplication rate of 99.7\% was identified. These duplicates were explicitly dropped rather than retained. While they represent realistic temporal patterns in CAN bus communications, where ECUs broadcast identical messages at fixed intervals, they also affect model training for overfitting and data leakage.

\begin{table}[htbp!]
    \centering
    \caption{Duplicate Analysis by Category}
    \resizebox{\columnwidth}{!}{
        \begin{tabular}{|l|c|c|c|c|}
            \hline
            \textbf{Category} & \textbf{Duplicates} & \textbf{Total Records} & \textbf{Percentage} & \textbf{Unique Messages} \\
            \hline
            BENIGN & 1,220,190 & 1,223,737 & 99.7\% & 3,547 \\
            DoS & 74,642 & 74,663 & 100.0\% & 21 \\
            SPOOFING & 109,799 & 109,819 & 100.0\% & 20 \\
            \hline
        \end{tabular}
    }
    \label{tab:duplicate_analysis}
\end{table}

\textit{Standardization}: To address inconsistent formatting, string manipulation was applied to the label, category, and specific\_class columns to strip whitespace and convert values to uppercase. Additionally, the DATA\_0 through DATA\_7 columns were checked and stripped of whitespace if they contained string objects.

\subsection{Exploratory Data Analysis}
The EDA phase focused on understanding class distributions, feature relationships, and attack signatures using Visualization techniques.

\textit{Class Distribution:} TABLE \ref{tab:basic_statistics} and \ref{tab:specific_class_dist} showing, the dataset exhibits a significant class imbalance typical of real-world IoT traffic, with Benign traffic dominating at 86.90\% (1,223,737 records) and Malicious traffic comprising 13.10\% (184,482 records).

\begin{table}[htbp]
    \centering
    \caption{Specific Class Distribution}
    \begin{tabular}{|l|c|c|}
        \hline
        \textbf{Specific Class} & \textbf{Count} & \textbf{Percentage} \\
        \hline
        BENIGN & 1,223,737 & 86.90\% \\
        DoS & 74,663 & 5.30\% \\
        RPM & 54,900 & 3.90\% \\
        SPEED & 24,951 & 1.80\% \\
        STEERING\_WHEEL & 19,977 & 1.42\% \\
        GAS & 9,991 & 0.71\% \\
        \hline
    \end{tabular}
    \label{tab:specific_class_dist}
\end{table}

\textit{DoS Attacks:} These represent the largest attack category with 74,663 messages (5.3\%) Table \ref{tab:specific_class_dist}.

\textit{Spoofing Attacks:} These account for 7.8\% of the data, with RPM spoofing being the most prevalent variant (54,900 messages) and GAS spoofing being the rarest (9,991 messages) Table \ref{tab:specific_class_dist}.

\textit{Feature Analysis (CAN IDs and Payloads)}: Visualizations of CAN IDs revealed distinct separation between attack types. * DoS Signature: Denial of Service attacks exclusively target CAN ID 291 (100\% concentration), making this feature a primary detector for DoS.

\begin{figure}[htbp!]
    \centering
    \includegraphics[width=1\linewidth]{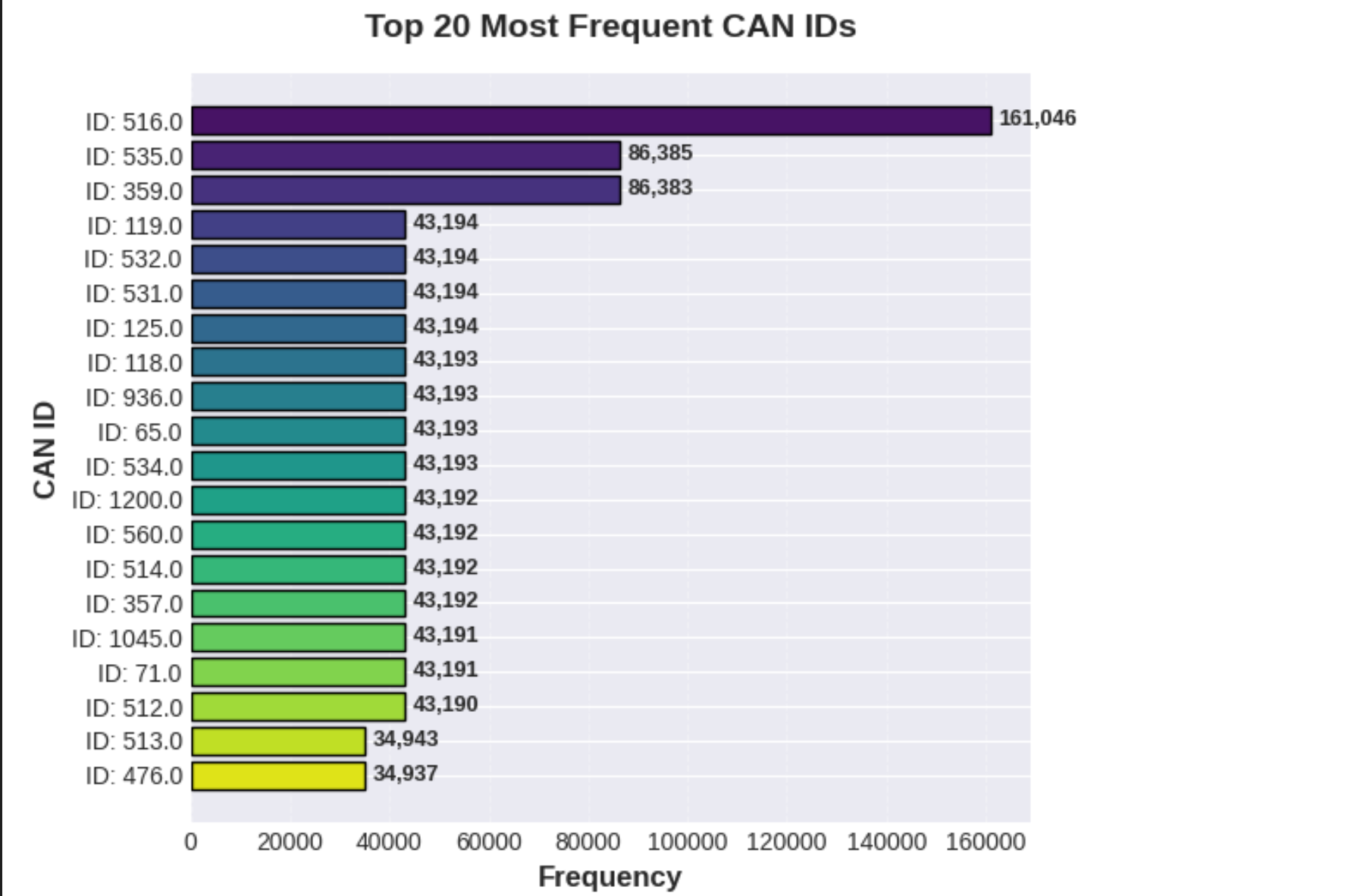}
    \caption{Top 20 Most Frequent CAN IDs}
    \label{fig:Top 20}
\end{figure}

\textit{Spoofing Signature:} Spoofing attacks focus on four primary IDs: 513, 476, 128, and 344 as showed in Figure \ref{fig:Top 20}. \

\textit{Benign Signature:} Normal traffic utilizes a diverse range of IDs, with 535, 516, and 359 being the most frequent as showed in Figure \ref{fig:Top 20}.

\begin{figure}[htbp!]
    \centering
    \includegraphics[width=1\linewidth]{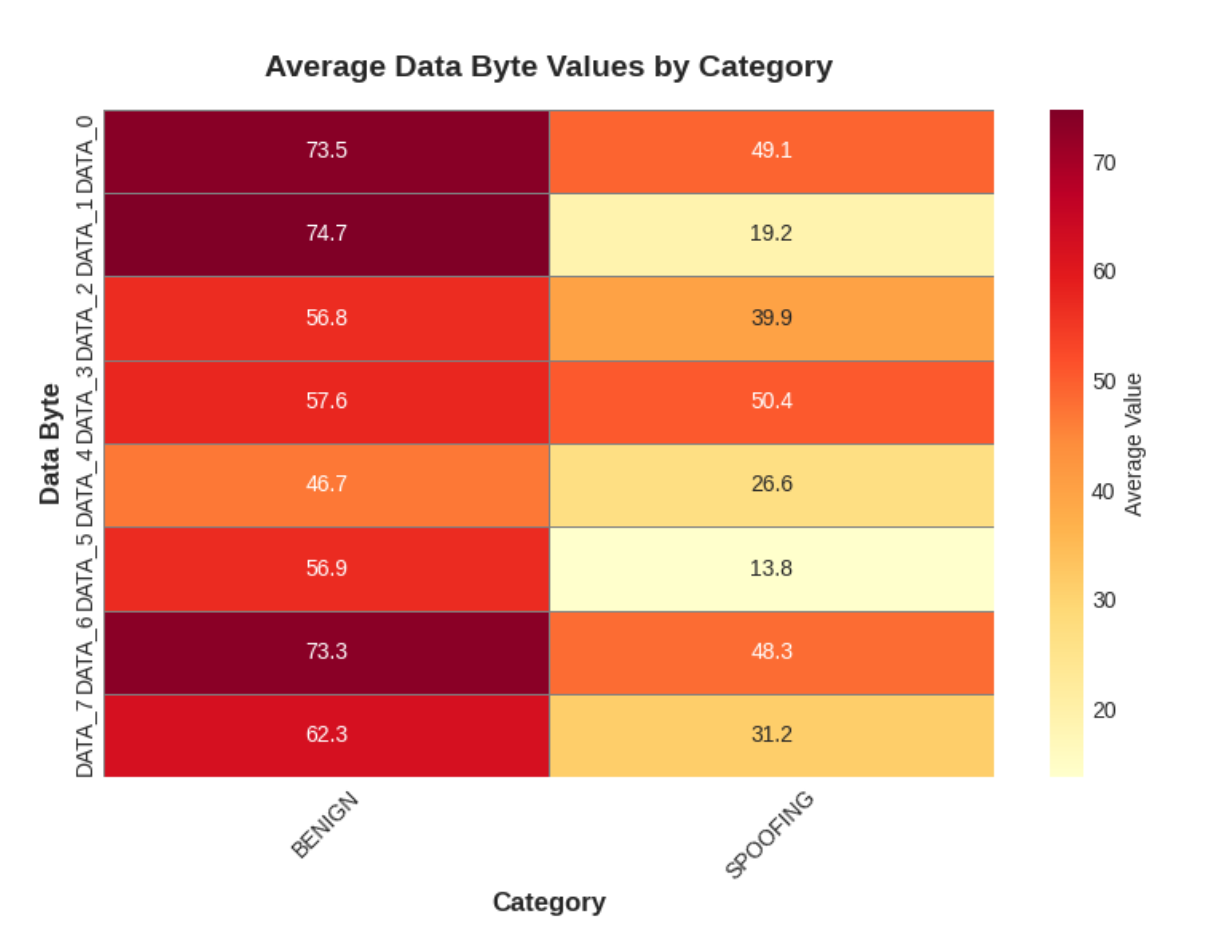}
    \caption{Average Data Byte Values By Category}
    \label{fig:Average Data}
\end{figure}

\begin{figure}
    \centering
    \includegraphics[width=1\linewidth]{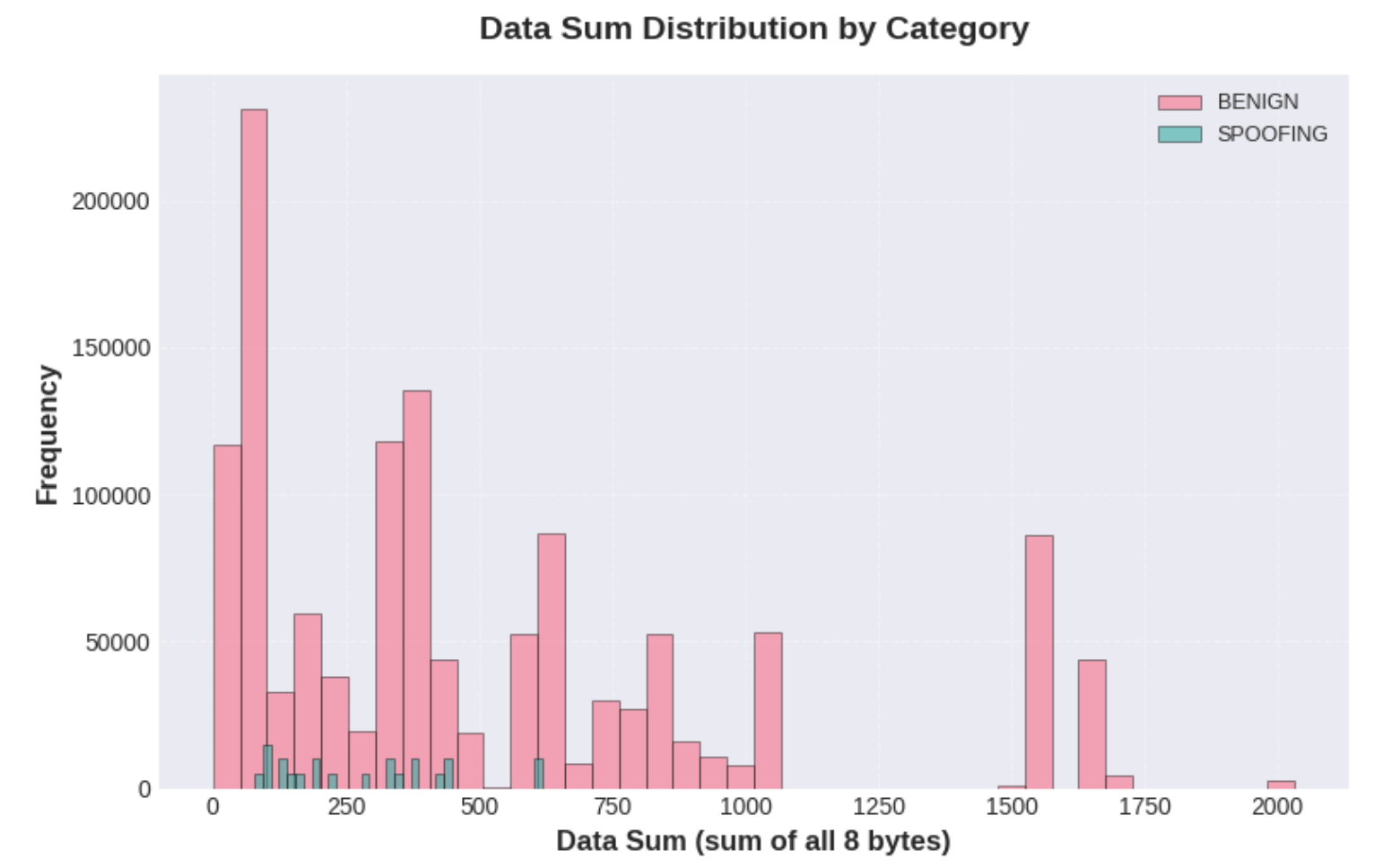}
    \caption{Data Sum Distribution by Category}
    \label{fig:Data Sum}
\end{figure}

\textit{Data Bytes:} Heatmaps Figure \ref{fig:Average Data} and histograms Figure \ref{fig:Data Sum} of the data bytes (DATA\_0 to DATA\_7) confirmed statistically significant differences between benign and malicious payload patterns.


\textit{Correlation Analysis:} A correlation matrix was generated to examine relationships between the 12 original features and a binary is\_attack target.

\textit{ID Correlation:} There is a weak negative correlation (-0.22) between ID and is\_attack, suggesting specific ID ranges are associated with attacks.

\begin{figure}[htbp!]
    \centering
    \includegraphics[width=0.75\linewidth]{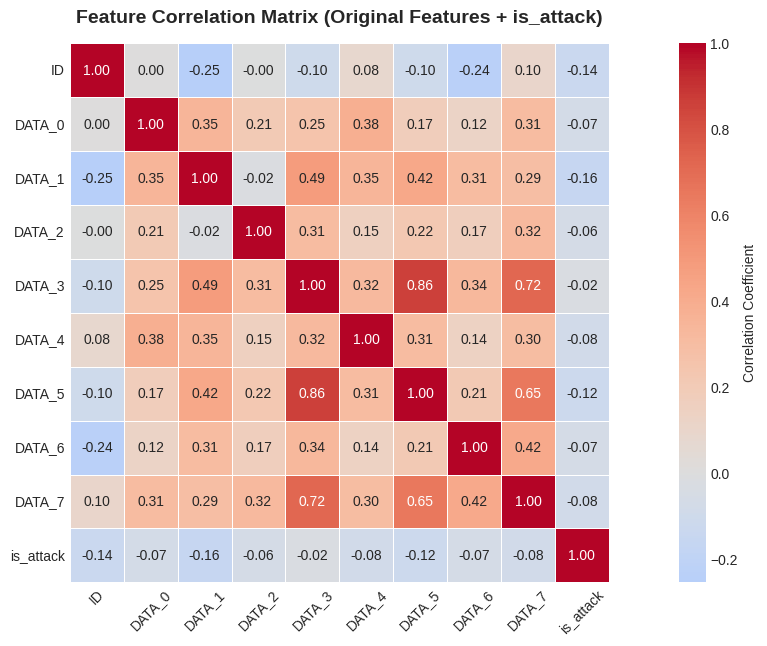}
    \caption{Feature Correlation Matrix}
    \label{fig:Feature Corr}
\end{figure}

\textit{Multicollinearity:} Figure \ref{fig:Feature Corr} shows High correlations were observed between specific data bytes, such as DATA\_3 and DATA\_5 (0.86) and DATA\_3 and DATA\_7 (0.72), indicating these bytes likely represent related vehicle parameters .

\subsection{Feature Selection Engineering}
High\-dimensional data often introduces noise and computational inefficiency, leading to the issues of dimensionality. Feature Selection engineering is a critical step in the machine learning pipeline to mitigate these issues. This report compares three standard techniques:\\

\subsubsection{Principal Component Analysis (PCA)}
PCA is utilized to perform dimensionality reduction by projecting data points onto new orthogonal axes (Principal Components) that maximize variance. As an unsupervised method, it operates without knowledge of target class labels. The data was scaled prior to application.

\begin{figure}[htbp!]
    \centering
    \includegraphics[width=1\linewidth]{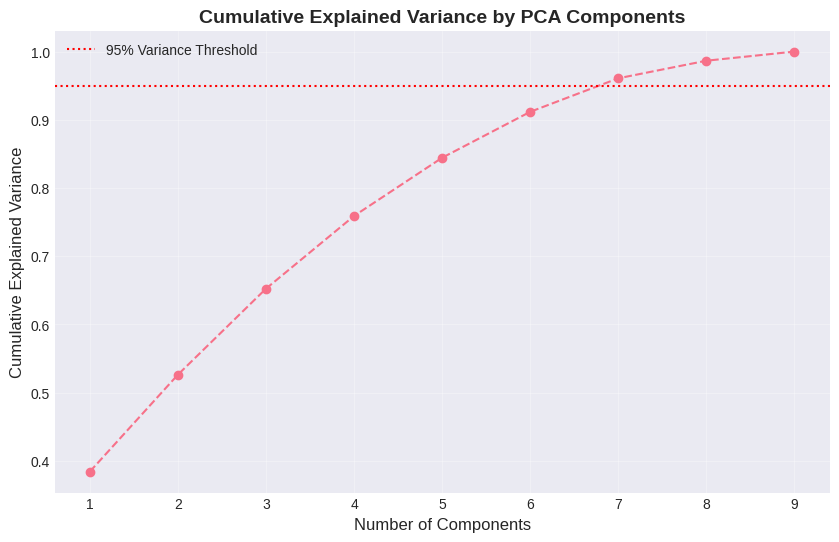}
    \caption{Cumulative Explained Variance PCA Components}
    \label{fig:PCA}
\end{figure}

\subsubsection{Linear Discriminant Analysis (LDA)}
LDA serves as both a dimensionality reduction and a classification technique. Unlike PCA, LDA is supervised; it computes new axes that maximize the distance between class means while minimizing the variance within each class (maximizing the Fisher criterion).

\begin{figure}[htbp!]
    \centering
    \includegraphics[width=1\linewidth]{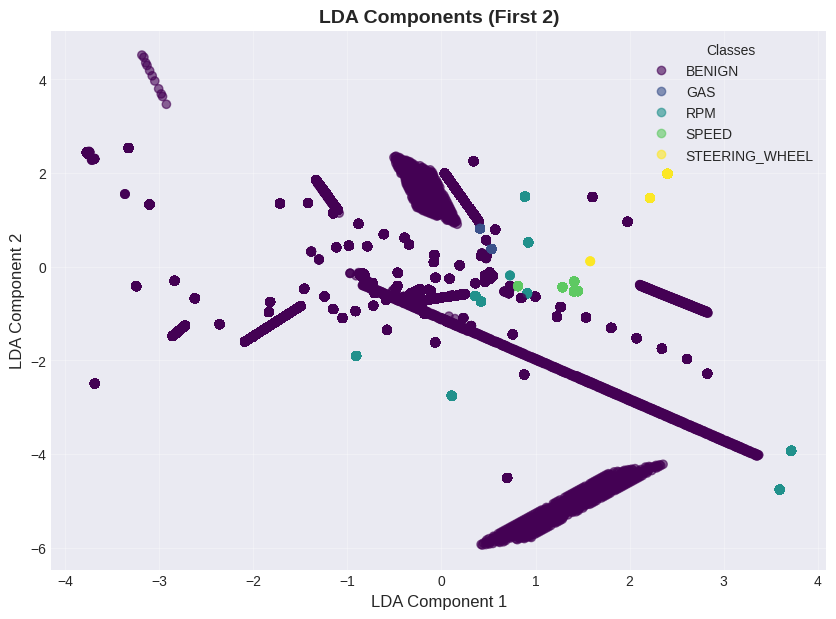}
    \caption{LDA Components}
    \label{fig:LDA}
\end{figure}

\subsubsection{ANOVA F-value}
Selection Analysis of Variance (ANOVA) is applied as a filter method. It evaluates each feature individually, calculating the F-value to determine if the means of the different classes are significantly different. Features with the highest F-scores are selected.\\

\begin{figure}[htbp!]
    \centering
    \includegraphics[width=1\linewidth]{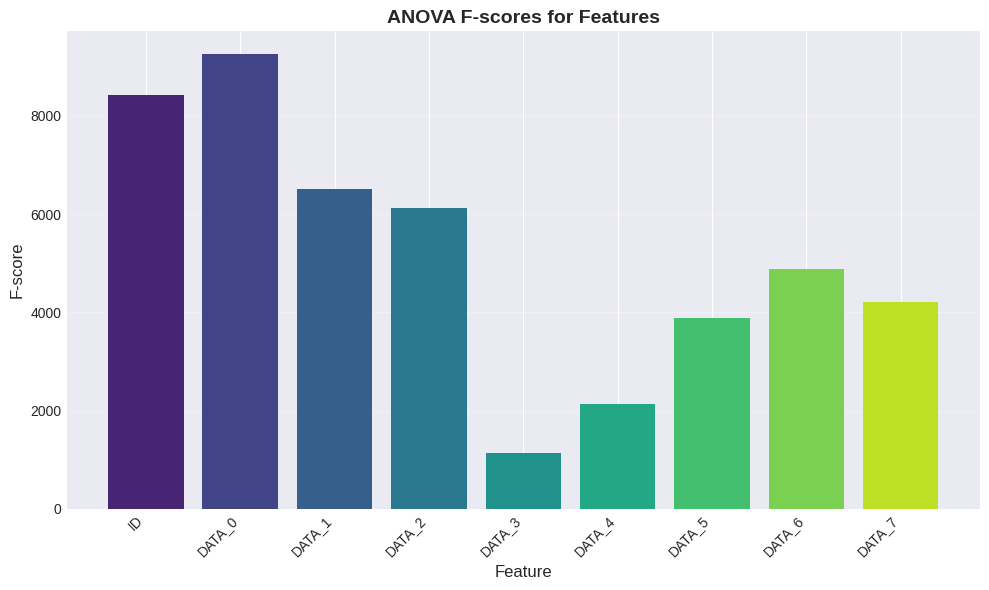}
    \caption{ANOVA F-scores for Features}
    \label{fig:ANOVA}
\end{figure}

The goal is to determine the optimal method for a dataset with $N=9$ features to prepare it for a subsequent classification algorithm.

PCA successfully reduced the dimensionality from 9 features to 7 principal components.Variance Retained: The 7 components retained approximately 95\% of the original information.Observation: While effective for noise reduction, the new features are linear combinations of the old ones, making semantic interpretation difficult.
The variance\-based approach did not explicitly account for class groupings.

LDA reduced the dimensionality from 9 features to 4 components (bounded by $C-1$ classes).Separability: Visual inspection of the projected space showed distinct, well-separated class clusters.

\textit{Efficiency:} LDA achieved better separation with fewer dimensions (4) compared to PCA (7).
\textit{Limitations:} The method required assumptions of normal distribution and equal covariance matrices, and was sensitive to outliers.

ANOVA identified the top 5 original features with the strongest statistical relationship to the target.Selected Features: ['ID', 'DATA\_0', 'DATA\_1', 'DATA\_2', 'DATA\_6']. Unlike PCA and LDA, ANOVA retained the original feature names, providing clear insight into which specific data points (e.g., DATA\_0) drive class differences.

The choice of method implies a trade-off between interpretability and discriminative power. PCA is neutral; it effectively compresses data but does not guarantee that the retained variance is useful for classification. ANOVA is excellent for Why It reveals that features like DATA\_0 differ significantly between groups. However, it ignores feature interactions. LDA is the strongest performer for the specific goal of classification. By explicitly optimizing for class separation, it created a feature space where the model is least likely to make errors, despite the features being abstract.

\subsection{Models Training and Evaluation}
This phase of the project focused on developing, training, and evaluating machine learning models to detect intrusions in the Internet of Vehicles (IoV) network using the CICIoV2024 dataset. Building upon the exploratory analysis and initial pre-processing steps, this stage involved the comparative analysis of classical machine learning algorithms, deep learning architectures, and hybrid ensemble methods.\\

\subsubsection{Models Training}
A critical observation from the initial data analysis was the presence of massive redundancy, characteristic of CAN bus traffic where messages are broadcast cyclically.
To prevent model overfitting and data leakage, duplicate rows were removed.

\begin{itemize}
    \item \textit{Original Shape:} 1,408,219 rows
    \item \textit{Unique Shape:} 3,568 rows
    \item \textit{Reduction:} 99.75\%
\end{itemize}

This transformation converted the problem from a large-scale data task to a high-precision small-data classification task. Then, the unique dataset was split into training (70\%) and testing (30\%) sets using stratified sampling to maintain the distribution of rare attack classes (GAS, RPM, SPEED, STEERING\_WHEEL).

\begin{itemize}
    \item \textit{Training Samples:} 2,497
    \item \textit{Testing Samples:} 1,071
\end{itemize}

Feature Selection and Dimensionality Reduction
To optimize model performance and computational efficiency, three distinct dimensionality reduction techniques were applied and evaluated against the original feature set:

\begin{itemize}
    \item \textit{Principal Component Analysis (PCA):} Retained 95\% variance, resulting in 6-7 components.
    \item \textit{Linear Discriminant Analysis (LDA): }Projected data to maximize class separability, reducing features to 4 components.
    \item \textit{ANOVA F-Value Selection:} Selected the top 5 features statistically most relevant to the target variable (ID, DATA\_0, DATA\_1, DATA\_2, DATA\_6).
\end{itemize}

\textbf{Comparative Results:} Testing with Logistic Regression, Decision Trees, and Random Forest revealed that Original Features and ANOVA selection generally yielded the highest F1-Macro scores. PCA resulted in a significant drop in F1-Macro scores for tree-based models, suggesting that the principal components obscured critical non-linear boundaries required to detect specific spoofing attacks.

\subsubsection{Models Evaluation}
Baseline models were trained on the unique dataset. Due to the extreme class imbalance (benign traffic dominating), F1-Macro was utilized as the primary performance metric alongside Accuracy.

\textit{Random Forest \& Decision Tree:} Achieved near-perfect performance (Accuracy ~99.8\%, F1-Macro ~0.82). These models proved robust in handling the categorical nature of the CAN IDs and data bytes.

\textit{Logistic Regression:} Performed poorly on the F1-Macro metric (approx. 0.20), indicating it correctly classified the majority class (Benign) but failed significantly on minority attack classes due to the non-linear nature of the attack signatures.

\begin{table}[htbp!]
    \centering
    \caption{Model Performance by Feature Set}
    \resizebox{\columnwidth}{!}{
        \begin{tabular}{|l|l|c|c|c|}
            \hline
            \textbf{Feature Set} & \textbf{Model} & \textbf{Accuracy} & \textbf{F1 (Macro)} & \textbf{Time (s)} \\
            \hline
            \multirow{3}{*}{Original} 
              & Decision Tree & 0.9972 & 0.7931 & 0.0022 \\
              & Logistic Regression & 0.9944 & 0.1994 & 0.0071 \\
              & Random Forest & 0.9981 & 0.7598 & 0.0487 \\
            \hline
            \multirow{3}{*}{ANOVA} 
              & Decision Tree & 0.9981 & 0.7599 & 0.0014 \\
              & Logistic Regression & 0.9944 & 0.1994 & 0.0061 \\
              & Random Forest & 0.9981 & 0.8266 & 0.0447 \\
            \hline
            \multirow{3}{*}{LDA} 
              & Decision Tree & 0.9907 & 0.6354 & 0.0061 \\
              & Logistic Regression & 0.9944 & 0.1994 & 0.0079 \\
              & Random Forest & 0.9963 & 0.4996 & 0.1100 \\
            \hline
            \multirow{3}{*}{PCA} 
              & Decision Tree & 0.9888 & 0.5132 & 0.0090 \\
              & Logistic Regression & 0.9944 & 0.1994 & 0.0057 \\
              & Random Forest & 0.9963 & 0.3596 & 0.0983 \\
            \hline
        \end{tabular}
    }
    \label{tab:model_performance}
\end{table}
\begin{figure}
    \centering
    \includegraphics[width=1\linewidth]{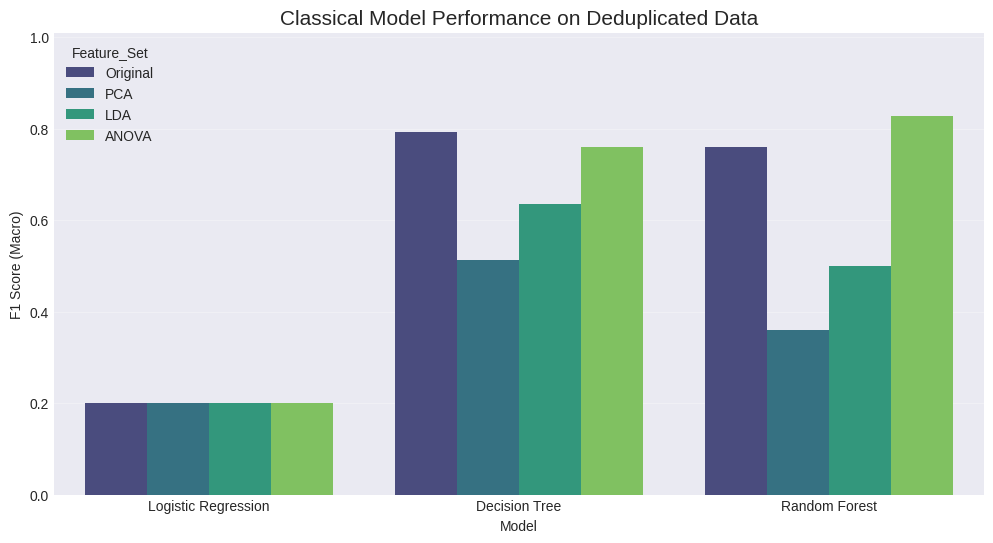}
    \caption{Classical Model Performance on Dedublicated Data}
    \label{fig:Class_model_performace}
\end{figure}
\textit{Deep Learning Models}: Two architectures were evaluated: a Deep Neural Network (MLP) and a 1D Convolutional Neural Network (CNN).
\begin{itemize}
    \item DNN (MLP): Accuracy: 99.63\%, F1-Macro: 0.6131
\begin{table}[htbp]
    \centering
    \caption{DNN MLP Results \& Execution Metrics}
    \begin{tabular}{|l|c|c|c|c|}
        \hline
        \textbf{Class} & \textbf{Precision} & \textbf{Recall} & \textbf{F1-Score} & \textbf{Support} \\
        \hline
        Class 0 & 0.9981 & 0.9991 & 0.9986 & 1065 \\
        Class 1 & 0.0000 & 0.0000 & 0.0000 & 1 \\
        Class 2 & 0.5000 & 0.3333 & 0.4000 & 3 \\
        Class 3 & 0.5000 & 1.0000 & 0.6667 & 1 \\
        Class 4 & 1.0000 & 1.0000 & 1.0000 & 1 \\
        \hline
        \multicolumn{5}{|c|}{\textbf{Weighted Averages}} \\
        \hline
        Macro Avg & 0.5996 & 0.6665 & 0.6131 & 1071 \\
        Weighted Avg & 0.9953 & 0.9963 & 0.9957 & 1071 \\
        \hline
        \multicolumn{5}{|c|}{\textbf{Overall Performance}} \\
        \hline
        \multicolumn{2}{|l|}{\textbf{Accuracy}} & \multicolumn{3}{c|}{99.63\%} \\
        \multicolumn{2}{|l|}{\textbf{F1-Macro}} & \multicolumn{3}{c|}{0.6131} \\
        \hline
    \end{tabular}
    \label{tab:DNN MLP report}
\end{table}
    \item 1D CNN: Accuracy: 99.53\%, F1-Macro: 0.2995, TABLE \ref{tab:1d_cnn_results}
    \begin{table}[htbp]
    \centering
    \caption{1D CNN Results \& Execution Metrics}
    \resizebox{\columnwidth}{!}{
        \begin{tabular}{|l|c|c|c|c|}
            \hline
            \textbf{Class} & \textbf{Precision} & \textbf{Recall} & \textbf{F1-Score} & \textbf{Support} \\
            \hline
            Class 0 & 0.9953 & 1.0000 & 0.9977 & 1065 \\
            Class 1 & 0.0000 & 0.0000 & 0.0000 & 1 \\
            Class 2 & 1.0000 & 0.3333 & 0.5000 & 3 \\
            Class 3 & 0.0000 & 0.0000 & 0.0000 & 1 \\
            Class 4 & 0.0000 & 0.0000 & 0.0000 & 1 \\
            \hline
            \multicolumn{5}{|c|}{\textbf{Weighted Averages}} \\
            \hline
            Macro Avg & 0.3991 & 0.2667 & 0.2995 & 1071 \\
            Weighted Avg & 0.9926 & 0.9953 & 0.9935 & 1071 \\
            \hline
            \multicolumn{5}{|c|}{\textbf{Overall Performance}} \\
            \hline
            \multicolumn{2}{|l|}{\textbf{Accuracy}} & \multicolumn{3}{c|}{99.53\%} \\
            \multicolumn{2}{|l|}{\textbf{F1-Score (Macro)}} & \multicolumn{3}{c|}{0.2995} \\
            \hline
        \end{tabular}
    }
    \label{tab:1d_cnn_results}
\end{table}
\end{itemize}

Despite high accuracy, the deep learning models struggled with the minority classes (F1-Macro score drop). The drastic reduction in data volume (down to ~2,500 training samples) meant there was insufficient data for these complex architectures to generalize distinct features for the rare attack types (e.g., only 1 test sample for GAS or SPEED attacks).

\textbf{Hybrid Ensemble Strategy}: To maximize detection rates, a custom ensemble was constructed using a combination of:

\begin{itemize}
    \item K-Nearest Neighbors (KNN) (Distance-weighted)
    \item Decision Trees (Gini \& Entropy variations)
    \item Support Vector Machine (SVM) (RBF Kernel)
    \item Multi-layer Perceptron (MLP)
\end{itemize}

A Hybrid Consensus approach was implemented. The system computed both a Hard Vote (majority rule) and a Weighted Soft Vote (probability-based). If both methods agreed, the prediction was kept; if they disagreed, the Soft Vote was preferred.

\begin{table}[htbp!]
    \centering
    \caption{Final Ensemble Model Results}
    \begin{tabular}{|l|c|c|c|c|}
        \hline
        \textbf{Class} & \textbf{Precision} & \textbf{Recall} & \textbf{F1-Score} & \textbf{Support} \\
        \hline
        BENIGN & 1.00 & 1.00 & 1.00 & 1065 \\
        GAS & 1.00 & 1.00 & 1.00 & 1 \\
        RPM & 1.00 & 0.67 & 0.80 & 3 \\
        SPEED & 1.00 & 1.00 & 1.00 & 1 \\
        STEERING & 0.00 & 0.00 & 0.00 & 1 \\
        \hline
        \multicolumn{5}{|c|}{\textbf{Overall Performance}} \\
        \hline
        \multicolumn{2}{|l|}{\textbf{Accuracy}} & \multicolumn{3}{c|}{99.81\%} \\
        \multicolumn{2}{|l|}{\textbf{Macro F1-Score}} & \multicolumn{3}{c|}{0.7598} \\
        \hline
    \end{tabular}
    \label{tab:ensemble_results}
\end{table}

\subsection{Key Findings}

\begin{table}[htbp]
    \centering
    \caption{Comprehensive Performance Comparison}
    \resizebox{\columnwidth}{!}{
        \begin{tabular}{|l|l|c|c|}
            \hline
            \textbf{Feature Set} & \textbf{Model} & \textbf{Accuracy} & \textbf{F1-Macro} \\
            \hline
            \multirow{3}{*}{ANOVA} 
             & Logistic Regression & 0.9944 & 0.1994 \\
             & Decision Tree & 0.9981 & 0.7599 \\
             & Random Forest & 0.9981 & 0.8266 \\
            \hline
            \multirow{3}{*}{LDA} 
             & Logistic Regression & 0.9944 & 0.1994 \\
             & Decision Tree & 0.9907 & 0.6354 \\
             & Random Forest & 0.9963 & 0.4996 \\
            \hline
            \multirow{3}{*}{PCA} 
             & Logistic Regression & 0.9944 & 0.1994 \\
             & Decision Tree & 0.9888 & 0.5132 \\
             & Random Forest & 0.9963 & 0.3596 \\
            \hline
            \multirow{5}{*}{Original} 
             & Logistic Regression & 0.9944 & 0.1994 \\
             & Decision Tree & 0.9972 & 0.7931 \\
             & Random Forest & 0.9981 & 0.7598 \\
             & 1D CNN & 0.9953 & 0.2995 \\
             & DNN MLP & 0.9963 & 0.6131 \\
            \hline
            \multirow{3}{*}{\shortstack{Original\\Ensemble}} 
             & Manual Hard Voting & 0.9981 & 0.7598 \\
             & Manual Soft Voting & 0.9981 & 0.7598 \\
             & Final Hybrid Ensemble & 0.9981 & 0.7598 \\
            \hline
        \end{tabular}
        }
    \label{tab:comprehensive_results}
\end{table}
\textit{The Accuracy Trap:} Across all models, accuracy remained consistently high (higher than 99\%) because the test set was dominated by Benign traffic (1065 out of 1071 samples). However, F1-Macro scores revealed that many models, particularly Logistic Regression and CNNs, completely ignored minority attack classes.

\textit{Feature Importance:} The specific CAN ID and the first three data bytes (DATA\_0 to DATA\_2) were identified via ANOVA as the most critical discriminators for identifying spoofing attacks.

\textit{Model Superiority:} The K-Nearest Neighbors (KNN) and Random Forest algorithms outperformed deep learning approaches on this reduced dataset. They successfully captured local patterns in the feature space that defined the rare attacks.

\textit{Ensemble Limitations:} While the final hybrid ensemble performed well, it failed to detect the Steering\_Wheel attack class (0 recall). This indicates that despite stratification, the extreme rarity of specific attack samples requires either synthetic data generation (SMOTE) or anomaly-detection-based approaches rather than standard supervised classification for future iterations.

\section{A comparison study with Kaggle works in Section~\ref{sec:Related Work}}
The findings from this project align with and expand upon recent literature regarding the CICIoT2023\cite{k2} and CICIoV2024\cite{k7} datasets.

\textit{The "Perfect Score"\cite{r7} Reality Check:} Literature utilizing advanced sampling techniques like Hybrid Sampling or Random Under-Sampling (RUS) often reports perfect 1.0 metrics for spoofing attacks. In contrast, this study employed aggressive de-duplication rather than synthetic balancing. The resulting F1-Macro scores (0.76–0.82) contradict the "perfect score" narrative, validating the dataset creators' original claim that spoofing attacks are inherently difficult to classify. This suggests that the perfect scores in other studies may stem from data leakage or over-simplified decision boundaries caused by sampling redundant data.

\textit{Algorithm Efficacy:} Consistent with studies on CICIoT2023\cite{rk2}, Random Forest emerged as the dominant architecture, successfully managing feature complexity where linear models like Logistic Regression failed. This confirms that tree-based ensembles are currently the most reliable choice for tabular IoV intrusion data.

\textit{Deep Learning and Overfitting:} While some papers report 99.8\% accuracy with Deep Neural Networks (DNNs), others caution that such high metrics often signal memorization rather than generalization. The results here support this skepticism: while the MLP and CNN achieved higher than 99\% accuracy, their low F1-Macro scores on the de-duplicated validation set reveal a failure to learn generalized features for minority classes.

\textit{Feature Engineering:} Research has shown that dimensionality reduction (like GSK clustering) can enhance MLP performance. Similarly, this study found that feature selection was critical. However, unlike some findings where PCA enhanced efficiency, PCA here degraded performance for tree-based models, suggesting that the raw ID and DATA fields contain non-linear information vital for detection that linear projections destroy.

\subsection{Conclusion}
This paper confirms that Random Forest and KNN are very suitable models for this specific IoV dataset when rigorous pre-processing is applied. The study exposes the "Accuracy Trap," where high accuracy masks poor minority class detection. Future work must address the detection of rare attacks like Steering\_Wheel, likely through synthetic oversampling (SMOTE) rather than simple replication, to bridge the gap between high precision and broad recall.


\end{document}